\documentclass[aps,pre,twocolumn,superscriptaddress,showpacs,floatfix]{revtex4}
\usepackage{graphicx}
\usepackage{bm}
\usepackage{amsmath}
\usepackage{amssymb}
\usepackage{color} 
\usepackage{ulem}  
\usepackage{float}
\usepackage{multirow}  
\usepackage{array}     

\input{epsf}

\begin{document}

\title{Thermoelectricity of interacting particles: A numerical approach}
\author{Shunda Chen}
\affiliation{Center for Nonlinear and Complex Systems,
Universit\`a degli Studi dell'Insubria, via Valleggio 11, 22100 Como, Italy}
\affiliation{Istituto Nazionale di Fisica Nucleare, Sezione di Milano,
via Celoria 16, 20133 Milano, Italy}
\author{Jiao Wang}
\affiliation{Department of Physics and Institute of Theoretical Physics
and Astrophysics, Xiamen University, Xiamen 361005, Fujian, China}
\author{Giulio Casati}
\affiliation{Center for Nonlinear and Complex Systems,
Universit\`a degli Studi dell'Insubria, via Valleggio 11, 22100 Como, Italy}
\affiliation{International Institute of Physics, Federal University of
Rio Grande do Norte, Natal, Brazil}
\author{Giuliano Benenti}
\affiliation{Center for Nonlinear and Complex Systems,
Universit\`a degli Studi dell'Insubria, via Valleggio 11, 22100 Como, Italy}
\affiliation{Istituto Nazionale di Fisica Nucleare, Sezione di Milano,
via Celoria 16, 20133 Milano, Italy}

\date{\today}

\begin{abstract}
A method for computing the thermopower in interacting systems is proposed.
This approach, which relies on Monte Carlo simulations, is illustrated first for a diatomic chain of hard-point elastically colliding
particles and then in the case of a one-dimensional gas with (screened)
Coulomb interparticle interaction. Numerical simulations up to $N>10^4$
particles confirm the
general theoretical arguments for momentum-conserving systems and show that
the thermoelectric figure of merit increases linearly with the system size.
\end{abstract}


\maketitle

\section{Introduction}

Thermoelectric materials are of interest due to their ability to convert
waste heat into electricity by the Seebeck effect or use electricity for
cooling by the Peltier effect~\cite{dresselhaus, snyder, shakuori, dubi,
thermo_review}. The efficiency of a thermoelectric material is a
monotonously growing function of the dimensionless figure of merit,
\begin{equation} \label{ZT1}
ZT=\frac{\sigma S^2}{\kappa}\,T,
\end{equation}
where $T$ is the temperature, $\sigma$ is the electrical conductivity,
$\kappa$ is the thermal conductivity, and $S$ is the thermopower (or
Seebeck coefficient). Increasing $ZT$ is a challenging task due to the
interdependency of transport coefficients. In particular,
the electrical conductivity and the electronic contribution to the
thermal conductivity are related by the Wiedemann-Franz law~\cite{kittel},
which, follows at low enough temperatures from the single-particle Fermi
liquid theory, states that the ratio $\sigma T/\kappa$ is constant.
Such limitations could be overcome, in principle, by the energy filtering
mechanism~\cite{mahansofo,linke1,linke2}, i.e., when transmission of
electrons is possible only within a tiny energy window.

On the other hand, very little is known about the thermoelectric
properties of interacting systems. In this case, analytical results are
rare and numerical simulations face difficult problems. To numerically
evaluate $ZT$, one may put the system into contact with two thermochemical
baths (reservoirs) allowing for particle exchange with the system. The
reservoirs are tuned at different temperatures and electrochemical
potentials in order to maintain stationary particle and heat currents.
By analyzing the response of these currents to the temperature and
electrochemical potential difference, one can evaluate the transport
coefficients and, in turn, $ZT$, with Eq.~(\ref{ZT1}). This method has
been successfully applied to the one-dimensional (1D) dimerized gas of
interacting hard-point particles~\cite{Benenti}. In that model, due to
the fact that particles only interact via instantaneous collisions, the
particles in the reservoirs are in effect decoupled from those of the
system. As such the reservoirs can be modeled as ideal gases so that
the simulations can be facilitated greatly (see Ref.~\cite{YT} for
the detailed description of the algorithm). However, in more general
systems with realistic interaction, the coupling between the reservoirs
and the system is essential; the mere injection of particles from an
ideal gas into the system may induce huge, unphysical interaction
energy when an injected particle is too close to a system particle.

Given this difficulty, which is unsolved yet to the best of our
knowledge, we turn to the closed  
heat baths~\cite{Lepri03, Dharrev}
that only exchange heat with the system. We sandwich the system with
two closed heat baths at different temperatures to establish a
nonequilibrium setup. For interacting systems, the heat conductivity
can be computed
directly with this setup, but computing thermopower
is challenging. In this paper, we solve this problem by the following
steps: We first use the grand-canonical Monte Carlo method~\cite{DFrenkel}
to compute
the electrochemical potential, $\mu$, as a function of the
particle density, $\rho$, at a given temperature; then, with our
nonequilibrium setup and by molecular dynamics, we compute the density
difference $\Delta \rho$ across the system, set in response to the
temperature difference $\Delta T$ applied to the system. Finally, based
on the established relation between $\mu$ and $\rho$, we map the density
difference into the thermoelectric voltage difference, $\Delta V$, so
that the thermopower is computed as $S=-\Delta V/\Delta T$. As to the
electrical conductivity $\sigma$, due to the fact that closed heat
baths do not support a charge current, it cannot be computed
numerically with our nonequilibrium setup; However, it can be
computed in equilibrium simulations by using the Green-Kubo
formula~\cite{kubo}.

To test and illustrate the method for computing the thermopower in nonequlibrium simulations, we first consider the 1D dimerized hard-point gas model \cite{Casati86}. Then, we numerically investigate the
case of a 1D gas of particles with nearest-neighbor Coulomb interaction,
modeling a screened Coulomb interaction between electrons. Due to
momentum conservation (more generally, due to the existence of a single
relevant conserved quantity), we expect on general grounds~\cite{Benenti}
that the figure of merit $ZT$ diverges in the thermodynamic limit,
implying that the Carnot efficiency is reached in this limit. This
result, so far illustrated by means of toy models~\cite{Benenti,
Benenti14}, is here confirmed in a more realistic model. Moreover,
$ZT$ exhibits a rapid, \textit{linear} growth with the system size,
which is a consequence of the recently reported Fourier-like behavior
of thermal conductivity~\cite{Zhong12, chen, dhar, wang, savin,
Fourierlike}.

\section{Numerical method}

The equations connecting fluxes and thermodynamic forces within linear
irreversible thermodynamics are~\cite{callen, degrootmazur}
\begin{equation}
\left(
\begin{array}{c}
j_\rho\\
j_u
\end{array}
\right) = \left(
\begin{array}{cc}
L_{\rho \rho} & L_{\rho u} \\
L_{u \rho} & L_{u u}
\end{array}
\right) \left(
\begin{array}{c}
-\nabla(\beta\mu)\\
\nabla \beta
\end{array}
\right) ,
\label{eq:lresponse}
\end{equation}
where
$j_\rho$ is the local particle current, $j_u$ is the local energy
current, $\mu$ is the electrochemical potential, and $\beta=1/(k_B T)$
is the inverse temperature (we set the Boltzmann constant $k_B=1$). The
kinetic coefficients $L_{ij}$ (with $i,j=\rho,u$) are related to the
familiar transport coefficients as
\begin{equation} \label{transport}
\sigma=\frac{e^2}{T}\,L_{\rho\rho},
\quad\kappa=\frac{1}{T^2}\frac{\det\mathbb{L}}{L_{\rho\rho}},
\quad S=\frac{1}{eT}\left(\frac{L_{\rho u}}{L_{\rho\rho}}-\mu\right);
\end{equation}
here $e$ is the charge of each particle (set to be
$e=1$), and $\det\mathbb{L}$ denotes the determinant of the (Onsager) matrix
of kinetic coefficients. Thermodynamics imposes $\det \mathbb{L}\ge 0$,
$L_{\rho\rho}\ge 0$, and $L_{uu}\ge 0$, and the Onsager reciprocity
relations ensure that $L_{u\rho} =L_{\rho u}$. Following Eq.~(\ref{ZT1}),
the thermoelectric figure of merit thus reads
\begin{equation} \label{ZT}
ZT=\frac{(L_{u\rho}-\mu L_{\rho\rho})^2}{\det \mathbb{L}}.
\end{equation}
Hereafter we describe a method for the computation of the transport
coefficients, and consequently $ZT$, in a generic interacting system.
Note that, even though we consider the particle and energy flows along
one direction, the motion inside the system could be, in principle, two-
or three-dimensional.

\textit{Thermal conductivity.--}
We compute the thermal conductivity by nonequilibrium
simulations. Two 
heat baths~\cite{Lepri03, Dharrev} at temperatures
$T_L=T+\Delta T/2$ and $T_R=T-\Delta T/2$, respectively, are connected
to the two ends of the system. Then the system is evolved and after a long
enough relaxation stage, when the stationary state has been established,
the heat conductivity is evaluated as $\kappa=\overline{j_u}/(\Delta T/L)$,
where the overbar denotes time averaging and $L$ is the system size along
the direction of the heat flow. The distributions of the
temperature and the particle density are calculated at the stationary state
as well. The temperature is numerically computed as
$T(x)=2\epsilon_k(x)/\rho(x)$, where
$\rho(x)=\overline{\sum_i \delta(x-x_i)}$ is the particle density and
$\epsilon_k(x)= \overline{\sum_i \frac{m_i v_i^2}{2} \delta(x-x_i)}$
is the kinetic energy density \cite{Dhar01}.


\textit{Thermopower.--}
We use the nonequilibrium setup with heat baths described above and
prepare the system in the stationary state. The thermopower is defined
as the magnitude of the induced thermoelectric voltage in response to the
temperature difference across the system, i.e., $S\equiv-\Delta\mu/e\Delta
T$, where $\Delta \mu=\mu_L-\mu_R=e\Delta V$ is the induced electrochemical
potential difference. In our simulations, we first compute
the particle density $\rho_L$
and $\rho_R$ at the two ends of the system, and then map these values into
$\mu_L$ and $\mu_R$, respectively, by means of the grand-canonical Monte
Carlo method (see the Appendix~\ref{sec:MonteCarlo}).

\textit{Electrical conductivity.--}
As we consider closed heat baths that do not exchange particles with
the system, we cannot compute the electrical conductivity with our
nonequilibrium setup. For this purpose, one can turn to equilibrium
simulations by taking advantage of the Green-Kubo formula~\cite{kubo}.

\section{Numerical tests with the 1D dimerized gas model}

\begin{figure}[!]
\vskip-0.1cm
\hspace*{-0.1cm}
\includegraphics[width=9.1cm]{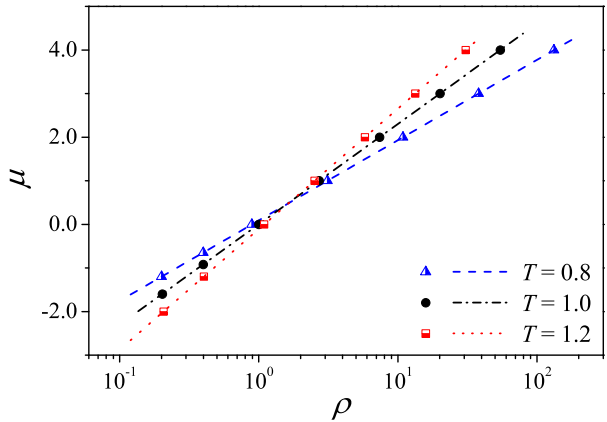}
\vskip-0.2cm
\caption{(Color online) Electrochemical potential versus particle density
for the equal-mass ($m=M$) hard-point gas. Numerical results of the
grand-canonical Monte Carlo (MC) simulations (symbols) are compared
with the analytical results (lines) given by Eq.~(\ref{eq:huang}).
Here and in the following figures we use units such that $m=1$, the
Boltzmann constant $k_B=1$, and the de Broglie thermal wavelength
$\lambda=1/\sqrt{T}$.}
\label{fig:montecarlo}
\end{figure}

In order to test and elucidate the grand-canonical Monte Carlo method
in computing the thermopower, we first consider a gas of $N$ colliding hard-point
particles with alternate masses $m$ and $M$, a paradigmatic model proposed
in Ref.~\cite{Casati86} and extensively investigated in the literatures
\cite{Dhar01, Savin02, Grassberger02, Casati03, Politi05, Dharrev, Hanggi11,
Chen13, SciChina, Chen14, Spohn14, Fourierlike, Benenti} for understanding
low-dimensional transport problem. We start with the case of equal masses,
$m=M$, for which the relation between the electrochemical potential and
the density is the same as for the one-dimensional ideal gas in the
semiclassical limit~\cite{huang}:
\begin{equation}
\mu=k_B T \ln(\rho\lambda).
\label{eq:huang}
\end{equation}
Here, $\lambda=h/\sqrt{2\pi mk_{B}T}$ is the de Broglie thermal wavelength
($h$ is the Planck's constant).  The excellent agreement between this
analytical expression and the numerical Monte Carlo simulations is shown
in Fig.~\ref{fig:montecarlo}. It is interesting to remark that from
classical thermodynamics of a one-dimensional ideal gas we obtain
${\mu}={k_B T}\ln({C \rho}/{\sqrt{T}})$, where the
constant $C$ cannot be determined by purely classical means. On the
other hand, such ambiguity is not present in the grand-canonical Monte
Carlo simulations. Indeed, this method is in some sense semi-classical,
in that it uses as information the value of the de Broglie thermal
wavelength and the grand canonical partition function where the particles
are considered as indistinguishable (see the Appendix~\ref{sec:MonteCarlo}; the factor $1/{N!}$ in
Eq.~(\ref{eq:grandcan}) is of purely quantum origin).
In our units, $\lambda = 1/\sqrt{T}$,  and therefore $C=1$.

\begin{table}[b]
\begin{tabular}{p{1cm}<{\centering}|
p{1cm}<{\centering}p{1cm}<{\centering}
p{1.1cm}<{\centering}p{1cm}<{\centering}
p{1cm}<{\centering}p{1cm}<{\centering}}  

\hline \hline

Size & $T'_L$ & $\rho_L$ & $\mu_L$ & $T'_R$ & $\rho_R$ & $\mu_R$ \\ \hline
21 &1.033& 0.978 & -0.040 & 0.979 & 1.022 & 0.032 \\  
41 & 1.037 & 0.971  &  -0.049  & 	0.967  & 1.031 & 0.046 \\   
81  & 1.041  & 0.965  & -0.058   & 0.960  & 1.037  & 0.054\\   
161  & 1.045  & 0.960  & -0.066  & 0.956  & 1.042  & 0.061 \\  
321  & 1.047  & 0.957  & -0.070  & 	 0.953  & 1.046  & 0.066 \\  
641  & 1.049  & 0.955  & -0.073  & 0.952  & 1.048  & 0.068 \\  
1281  & 1.049  & 0.954  & -0.074  & 0.951  & 1.050  & 0.070 \\  
2561  & 1.050  & 0.953  & -0.076  & 0.951  & 1.050 & 0.070 \\  
5121  & 1.050  & 0.953  & -0.076	& 0.950  & 1.051  & 0.072 \\  
10241  & 1.050  & 0.953	& -0.076	& 0.950	 & 1.051  & 0.072 \\ \hline \hline

\end{tabular}
\caption{The numerically computed data for hard-point gas model. The temperatures of the two heat baths are set to be $T_L=1.05$ and $T_R=0.95$.}
\label{table:detail}
\end{table}

\begin{figure}[t]
\vskip-0.8cm
\hspace*{-0.72cm}
\includegraphics[width=10.4cm]{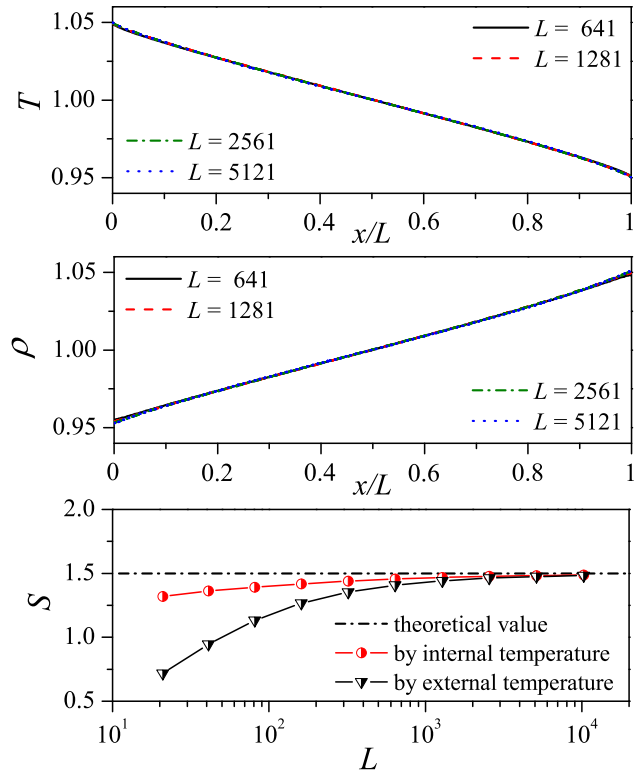}
\vskip-0.5cm
\caption{(Color online)
The temperature profile (top panel), the density profile (middle panel),
and the thermopower (bottom panel) for the hard-point gas with unequal
masses $m=1$ and $M=(\sqrt5+1)/2$. The temperatures of the two heat baths
are set to be $T_L=1.05$ and $T_R=0.95$. The dash-dotted line in the bottom
panel shows the analytical result of $S=3/2$.}
\vskip-0.5cm
\label{fig:seebeck}
\end{figure}

As shown in Fig.~\ref{fig:montecarlo}, with the grand-canonical
Monte Carlo method we can numerically determine the dependence of the
electrochemical potential on the particle density at a given temperature
$T$. Further, we can compute the thermopower by nonequilibrium  molecular
simulations using two statistical thermal baths,
with different temperatures $T_L$ and $T_R$, coupled to the left and the right end of the system. When the first (last) particle collides with the left (right) side of the system, it is injected back with a new speed $|v|$
determined by the distribution~\cite{heatbath}
\begin{equation}
P_{L,R}(v) = \frac{|v|m_{1,N}}{k_BT_{L,R}}\exp\left(-\frac{v^{2}m_{1,N}}{2k_BT_{L,R}}\right),
\end{equation}
where $m_{1}$ and $m_{N}$ are the masses of the first and the last particle.

As an example, here we consider the gas
of colliding hard-point particles with alternate masses $m=1$ and $M=(\sqrt5+1)/2\approx 1.618$ \cite{Fourierlike}. The mean distance between two nearest-neighboring particles is set to be unity, so that the system length (size) $L$ equals the particle number $N$. Figure~\ref{fig:seebeck} shows the stationary temperature and density profiles (top and middle panels).
The densities $\rho_L$ and $\rho_R$ at the left and right ends of the
chain can be computed directly. Then the relation between $\rho$ and
$\mu$ provided by the grand-canonical Monte Carlo simulations allows
us to obtain the corresponding values of $\mu_L$ and $\mu_R$ and to
compute the thermopower (bottom panel of Fig.~\ref{fig:seebeck}) as $S=-\Delta \mu/e\Delta T$. It should be noted that, as shown in Table~\ref{table:detail}, for small system sizes, the internal temperatures $T'_L$ and $T'_R$ at the left and right ends of the system, at which the particle densities $\rho_L$ and $\rho_R$ are computed, are slightly
different from the external temperatures $T_L$ and $T_R$.
This discontinuity is the result of a boundary resistance, generally
denoted as Kapitza resistance; see for instance Ref.~\cite{Lepri03}.
This boundary effect vanishes when increasing the system size;
see Table~\ref{table:detail}. However, it may be relevant for small systems sizes, as
shown in the bottom panel of Fig.~\ref{fig:seebeck}, where we compare
the thermopower computed by using the internal temperatures as
$S=-(\mu_L-\mu_R)/e(T'_L-T'_R)$ or
the external temperatures as $S=-(\mu_L-\mu_R)/e(T_L-T_R)$.
Since we are here interested in the intrinsic transport properties of the
system rather than in the details of the coupling to the heat baths,
in what follows we will show results for the first method only.
Moreover, we note that the first method shows a faster convergence
to the asymptotic analytical value $S=3/2$ \cite{Benenti} than the second one.
Finally, the convergence of our numerical results for a nonintegrable
model to the analytical value corroborates the validity of our
computational scheme.


\section{Thermoelectricity of the Coulomb gas}

With the help of the above described method, now we study a 1D system
of more general interaction.

\subsection{The model}

The model system we consider consists of $N$ charged particles with
nearest-neighbor Coulomb interaction, which is described by the
Hamiltonian
\begin{equation}
H=\sum_{i}\left[\frac{p_{i}^{2}}{2 m_i}+U(x_{i}-x_{i-1})\right],
\label{eq:hamiltonian}
\end{equation}
where $m_i$, $x_i$, and $p_i=m_i \dot{x}_i$ are the mass, the
coordinate, and the momentum of the $i$th particle, respectively,
and the nearest-neighbor interaction given by $U(x)=a/x$ can be considered
as a simplified effective model of screened interaction between particles
($a$ is the controlling parameter of interaction strength and,
due to Coulomb repulsion, particles cannot cross each other;
i.e., $x_{i}>x_{i-1}$ and $U(x)>0$). The overall
momentum $P=\sum m_{i} \dot{x}_{i}$ is conserved. The total charge current
is $J_e= e J_\rho$, where $J_{\rho}=\sum \dot{x}_{i}$ is the total particle
current. The total energy current reads as follows~\cite{Lepri03}: $J_{u}
=\sum j_{i}$, where the local energy current $j_{i}=\frac{1}{2}(x_{i+1}-
x_{i}) (\dot{x}_{i+1}+\dot{x}_{i})F(x_{i+1}-x_{i})+\dot{x}_{i}h_{i}$ with
$F(x)=-U'(x)$ and $h_{i}=[m_i \dot{x}^2_{i}+U(x_{i+1}-x_{i})+U(x_{i}-
x_{i-1})]/2$. In the following we assume that all particles have the same,
unitary mass; i.e., $m_i=m=1$. Moreover,
the mean distance between two nearest neighboring particles is set to be unity, so that the system length $L$ equals the particle number $N$.

\subsection{Simulation results}

\begin{figure}[!]
\vskip.2cm
\includegraphics[scale=0.82]{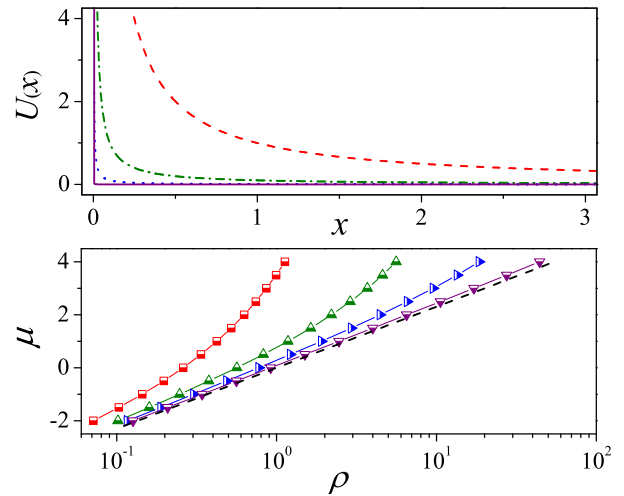}
\vskip-.0cm
\caption{Top: Coulomb potential $U(x)=a/x$ for various values of the
parameter $a$. Bottom: electrochemical potential versus particle density
at temperature $T=1$. In both panels, from top to bottom, $a=1, 10^{-1},
10^{-2}$, and $10^{-4}$, respectively. The straight dashed line shows the
analytical relation [Eq.~(\ref{eq:huang})] between $\mu$ and $\rho$ for
the hard-point gas.}
\label{fig:mccoulomb}
\end{figure}

We first compute by means of the grand-canonical Monte Carlo method
the mapping between the density and the electrochemical potential for
the Coulomb gas model. A broad value range of the interaction parameter $a$,
ranging from $a=10^{-4}$ to $a=1$, has been investigated. Note that, as
clearly shown in the top panel of Fig.~\ref{fig:mccoulomb}, in the limit
of $a\to 0$ the hard-point gas model is recovered. In addition, as expected
and shown in the bottom panel of Fig.~\ref{fig:mccoulomb}, the smaller
$a$ is, the closer the electrochemical potential (for a given particle
density) to the analytical result given by Eq.~(\ref{eq:huang})
predicted for the hard-point gas model.

For the computation of the thermopower and the thermal conductivity,
we use Langevin heat baths ~\cite{Dharrev} set at temperatures $T_L=T+\Delta T/2$ and
$T_R=T-\Delta T/2$. In our simulations, the value of $\Delta T$,
10\% of $T$, is set with the consideration that it is small enough
to guarantee the system to be in the linear response regime but meanwhile
not too small to facilitate the simulations. (Random tests with smaller
$\Delta T$, e.g., 4\% of $T$ have been done and the same results
have been obtained.) The system is evolved with velocity-Verlet algorithm,
but we have verified that all the results do not depend on the integration
algorithm. The relaxation stage is longer than $t=10^7$ for all the
simulated cases.


With regard to the electrical conductivity, in this case it is not
necessary to use the Green-Kubo formula. Indeed, thanks to momentum
conservation, $\sigma$ is ballistic and can be computed analytically.
Using a stochastic model of thermochemical baths~\cite{carlos2001,
carlos2003}, we have $j_\rho=\gamma_L-\gamma_R$, where $\gamma_\alpha=
\rho_\alpha \sqrt{k_B T}/\sqrt{2\pi m}$ is the injection rate of particles
from reservoir $\alpha$ into the system ($\alpha=L,R$ stands for the left
and right reservoir, respectively), and $\rho_\alpha$ is the density of
particles in reservoir $\alpha$, modeled as an infinite one-dimensional
ideal gas. Since the electrochemical potential $\mu_\alpha$ for reservoir
$\alpha$ is given by $\mu_\alpha=k_B T\ln(\lambda\rho_\alpha)$~\cite{huang},
we obtain $\gamma_\alpha=\frac{k_B T}{h} \exp(\beta\mu_\alpha)$ and, therefore,
within linear response regime, $j_\rho=\frac{\exp(\beta\mu_\alpha)}{h}\,
\Delta \mu$. The electrical conductivity is ballistic and given by
$\sigma=G L$, with the conductance
\begin{equation}
G=\frac{e j_\rho}{\Delta \mu}=\frac{e}{h}\,\exp(\beta\mu),
\label{eq:conductance}
\end{equation}
where, always within linear response, $\mu\approx \mu_L\approx \mu_R$.

\begin{figure}[!]
\vskip-0.95cm
\hspace*{-0.82cm}
\includegraphics[width=10.4cm]{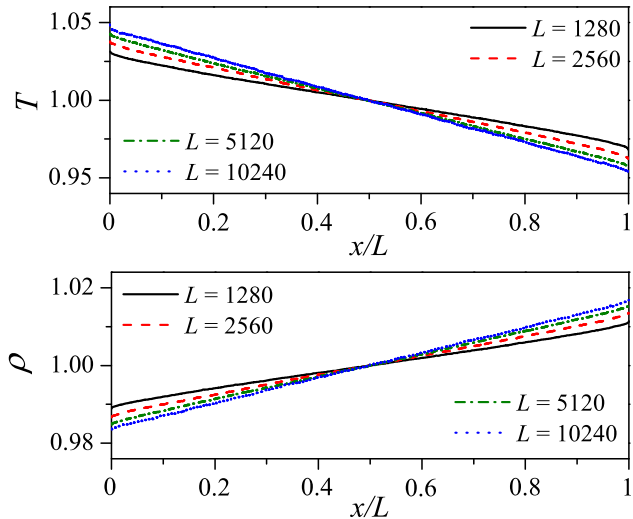}
\vskip-0.5cm

\caption{(Color online)
Temperature (top panel) and the density (bottom panel)
profiles for the Coulomb gas model. The temperatures of the two heat baths
are set to be $T_L=1.05$ and $T_R=0.95$.}
\label{fig:densitycoulomb}
\vskip-0.1cm
\end{figure}

\begin{figure}[!]
\vskip-.2cm
\hspace*{0.1cm}
\includegraphics[scale=1.15]{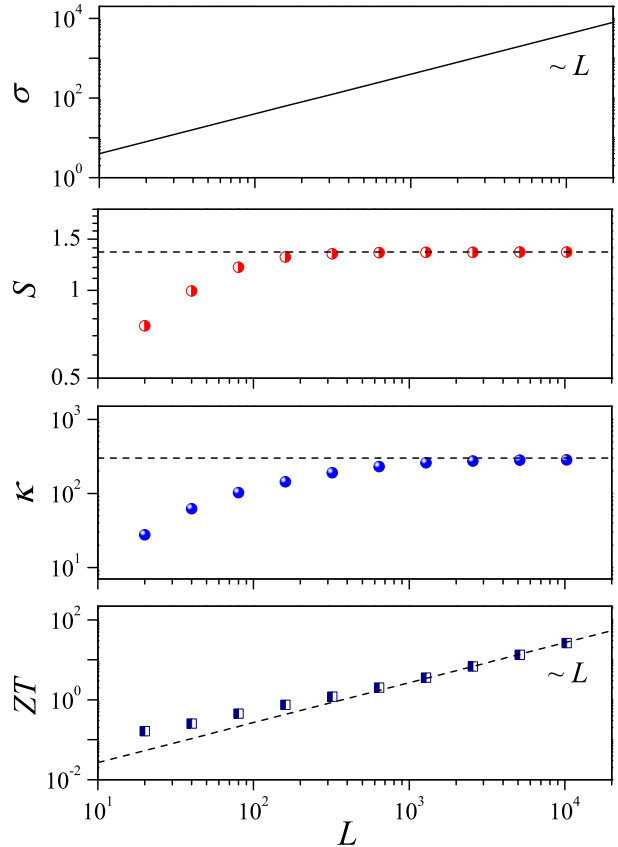}
\vskip-1.4cm
\caption{(Color online) Dependence of the transport coefficients $\sigma$,
$S$, $\kappa$ and $ZT$ on the system size $L$ for the Coulomb gas model.
Dashed lines are drawn for
reference. The temperature $T=1$ and the interaction strength $a=1$.}
\label{fig:nonequilibrium}
\vskip-0.2cm
\end{figure}

Figure~\ref{fig:densitycoulomb} shows the stationary temperature and density profiles for the Coulomb gas model. Similarly to the diatomic hard-point gas model, the system's temperatures at the boundaries approach the temperatures of the thermal baths when the system size $L\to\infty$; i.e., $T'_L\to T_L$ and
$T'_R\to T_R$. In Fig.~\ref{fig:nonequilibrium} we show the transport coefficients
$\sigma$, $S$, $\kappa$, and the thermoelectric figure of merit $ZT$.
For $\sigma$ we plot the analytically determined linear growth as a
function of the system size, $\sigma=G L$, with the conductance $G$
given by Eq.~(\ref{eq:conductance}) (we have checked that
consistent results,
i.e., $\sigma\sim L$, can also be obtained by the Green-Kubo formula
in equilibrium simulations,
where the integration time is correctly truncated to take into
account the ballistic transport~\cite{Lepri03}).
The thermal conductivity increases
for small system sizes and then saturates, as expected in a system that
obeys the Fourier law. This behavior has been reported in recent
investigations of several one-dimensional models of interacting
particles~\cite{Zhong12, chen, dhar, wang, savin, Fourierlike}. While
the Fourier-like regime might be an intermediate (in the system size)
regime, followed by an asymptotic regime of anomalous thermal conductivity
$\kappa\sim L^{1/3}$~\cite{Lepri03,Dharrev}, its range may expand rapidly
as an integrable limit (here, for $a\to 0$) is approached~\cite{Fourierlike}.
As a consequence, for practical purposes we can assume that the system
obeys the Fourier law. Since also the thermopower saturates as predicted
for ballistic transport [see Eq.~(\ref{eq:thermopower}) in
Sec.~\ref{sec:momentumconserving}], we can conclude that $ZT$ grows
linearly with the system size. In what follows, we show that the
divergence of $ZT$ with the system size can be explained in terms
of a general theoretical argument for momentum-conserving
systems~\cite{Benenti}.

\subsection{Thermoelectricity in momentum-conserving systems}
\label{sec:momentumconserving}

At the thermodynamic limit, the presence of nonzero Drude weights
${\cal D}_{ij}$ is a signature of ballistic transport~\cite{zotos, garst,
zotosreview, heidrich-meisner}; i.e., the kinetic coefficients $L_{ij}$
scale linearly with  the system size $L$. As a consequence, the thermopower
$S$ is asymptotically size-independent. The finite-size Drude weights,
for a system of size $L$, can be related to the existence of relevant
conserved quantities of the system and computed by means of the Suzuki
formula~\cite{Suzuki}. Such formula states that
\begin{equation}
\begin{array}{c}
{\displaystyle
C_{ij}(L)\equiv
\lim_{t\to\infty}\frac{1}{t}
\int_0^{t} dt' \langle J_i(t') J_j(0) \rangle_T
}
\\
\\
{\displaystyle
~~=
\sum_{n=1}^M
\frac{\langle J_i Q_n \rangle_T
\langle J_j Q_n \rangle_T}{\langle Q_n^2\rangle_T},
}
\end{array}
\label{eq:suzuki}
\end{equation}
where $\langle \cdots \rangle_T$ denotes the thermal average at
temperature $T$, and $\{Q_n$, $n=1,\cdots,M\}$ denote $M$ orthogonal
constants of motion, which are relevant; that is, non-orthogonal to the
considered currents, in our case to the currents $J_\rho$ and $J_u$:
$\langle J_\rho Q_n \rangle_T\ne 0$ and
$\langle J_u Q_n \rangle_T\ne 0$.
The finite-size Drude weights are then defined as
\begin{equation}
D_{ij}(L)\equiv \frac{1}{2 L}\,C_{ij}(L) \ .
 \label{drudeL}
\end{equation}
If the thermodynamic limit $L\to\infty$ commutes with the long-time
limit $t\to\infty$, then the thermodynamic Drude weights ${\cal D}_{ij}$
can be obtained as
\begin{equation}
{\cal D}_{ij}=\lim_{L\to\infty} D_{ij}(L).
\label{eq:drudeinfty}
\end{equation}
Moreover, if the limit does not vanish we can conclude that the presence
of relevant conservation laws yields nonzero generalized Drude weights,
which in turn implies ballistic transport.

We can see from Suzuki's formula that for systems with a single relevant
constant of motion ($M=1$), the ballistic contribution to $\det \mathbb{L}$
vanishes, since it is proportional to ${\cal D}_{\rho\rho}{\cal D}_{uu}-
{\cal D}_{\rho u}^2$, which is zero from Eqs. (\ref{eq:suzuki}),
(\ref{drudeL}), and
(\ref{eq:drudeinfty}). Hence, $\det \mathbb{L}$ grows slower than $L^2$,
and therefore the thermal conductivity $\kappa\sim \det{\mathbb{L}}/
L_{\rho\rho}$ grows sub-ballistically, $\kappa\sim L^\nu$, with $\nu<1$.
Furthermore, since $\sigma\sim L_{\rho\rho}\sim L$ is ballistic and
$S\sim L^0$, we can conclude that~\cite{Benenti}
\begin{equation}
ZT=\frac{\sigma S^2}{\kappa}\,T \propto  L^{1-\nu} \ .
\end{equation}
Hence, $ZT$ diverges in the thermodynamic limit $L\to\infty$. This general
theoretical argument applies, for instance, to systems where momentum is the
only relevant conserved quantity. It has so far been illustrated in a toy
model, i.e., a 1D dimerized gas of interacting hard-point
particles~\cite{Benenti}, and in a two-dimensional stochastic model of
interacting particles~\cite{Benenti14}. Here we consider the
more realistic and complex model of the Coulomb gas.

In order to check if our theory applies to the Coulomb gas model, in
particular if the thermodynamic limit $L\to\infty$ commutes with the
long-time limit $t\to\infty$, we need to compute the current correlation
functions. For this aim we perform the equilibrium simulations with periodic boundary conditions. We prepare the equilibrium state of the system by using
Andersen heat baths~\cite{Andersen} at the same given temperature $T$ and evolve the system for a sufficiently long time to make sure that it has been well thermalized. Then we remove the heat baths from the system. Starting from this moment the system is evolved isolatedly. After another long enough time of evolution, the correlation functions $\langle J_i(t) J_j(0)\rangle_T$ ($i,j=\rho,u$) are computed as functions of time.

\begin{figure}
\vskip-.35cm
\includegraphics[scale=1.16]{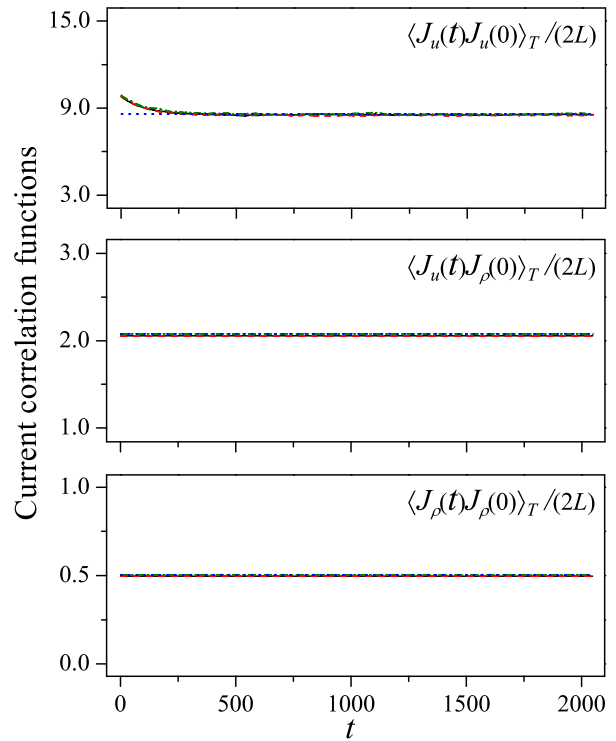}
\vskip-2.8cm
\caption{(Color online) Current-current correlation functions for
the Coulomb gas model at temperature $T=1$ and interaction strength $a=1$.
In all the panels, the black straight lines are for $L=64$, the red
dashed lines are for $L=128$, and green dash-dotted lines are for $L=256$.
The results of the finite-size Drude weights $D_{ij}(L)$ by Suzuki's formula
(blue dotted horizontal line for $L=256$) are also shown for comparison.}
\label{fig:correlations}
\end{figure}

The numerical results for the autocorrelation functions of $J_\rho$ and
$J_u$, and for the cross correlation function between them are shown in
Fig.~\ref{fig:correlations}. It can be seen that correlation function
$\langle J_u(t) J_u(0)\rangle_T$ approaches a finite nonzero value as the
correlation time increases and that the characteristic time scale to
approach such value is independent of the system size. Other
current-current correlation functions are constant. We have, therefore, a
strong numerical evidence that we can commute the long-time limit $t\to
\infty$ and the thermodynamic limit $L\to\infty$ and we can compute the
thermodynamic Drude weights ${\cal D}_{ij}$ by means of
Eq.~(\ref{eq:drudeinfty}). We also compute the finite-size Drude
weights $D_{ij}(L)$ via the Suzuki formula Eqs.~(\ref{eq:suzuki}) and
Eq.~(\ref{drudeL}), and indicate its value by a dotted horizontal line
in the plots of Fig.~\ref{fig:correlations}. (Note that it is not a
function of time $t$.) The obtained numerical results are in good agreement
with the (asymptotical) values of the correlation functions. We note that
$\langle J_\rho(t) J_\rho(0)\rangle_T $ does not decay; this is because
$J_\rho=P/m$ is a conserved quantity, so that the particle current is the
same as for the ideal gas, and therefore by employing the Suzuki's formula,
we can analytically obtain that $D_{\rho\rho}=T\rho/2$. This result is in
perfect agreement with the data shown in the last panel of
Fig.~\ref{fig:correlations}. In fact, $\langle J_\rho(t) J_u(0) \rangle_T
=P\langle J_u(0) \rangle_T/m$ is also trivially constant due to momentum
conservation. (See the middle panel of Fig.~\ref{fig:correlations}).
Finally, as expected from the theory~\cite{Benenti}, $D_{\rho\rho}(L)D_{uu}
(L)-D_{\rho u}^2(L)=0$; this is also verified for various system sizes.

For momentum-conserving systems, we can also compute the asymptotic
value of the thermopower (as $L\to\infty$) based on theoretical
prediction for ballistic transport:
\begin{equation}
S=\frac{1}{eT}\left(\frac{{\cal D}_{\rho u}}{{\cal D}_{\rho\rho}}
-c\right).
\label{eq:thermopower}
\end{equation}
Here, ${\cal D}_{ij}$ can be obtained with the above equilibrium simulations
via Eq.~(\ref{eq:drudeinfty}),
and $c$ is a numerical constant that can be determined by comparison
with the results obtained by the grand-canonical Monte Carlo
method~\cite{foot_mu}.

\section{Conclusions}

Starting from the definition of thermopower as a measure of the magnitude
of an induced thermoelectric voltage in response to a temperature difference
and taking advantage of the grand canonical Monte Carlo method to connect
the particle density to the electrochemical potential, we are able to compute
the thermoelectric coefficients in systems with more general interaction than
the instantaneous collisions. As a physically significant illustration of our
approach, we have shown that for classical one-dimensional, momentum-conserving
systems with (screened) Coulomb interaction, the thermoelectric figure of merit
increases, on a broad range of the system size, linearly. In principle, our
strategy can be applied without restrictions on the type of interaction,
even in the case of electron-lattice coupling, hence it could be useful
for studying thermoelectricity in more complex and realistic systems.

\begin{acknowledgments}
We acknowledge the support by the NSFC (Grants No. 11275159, No. 11535011, and No. 11335006),
by MIUR-PRIN, and by the CINECA project \textit{Nanostructures for Heat
Management and Thermoelectric Energy Conversion}.
\end{acknowledgments}

\appendix

\section{Grand-canonical Monte Carlo simulations}
\label{sec:MonteCarlo}

For a given electrochemical potential, system size, and temperature,
the grand-canonical Monte Carlo method samples the grand-canonical
probability distribution
\begin{equation}
f_{\mu L T}({\bf x}^{N}; N)\propto
\frac{L^{N}\exp(\beta\mu N)}{N!\,\lambda^{N}}\exp[-\beta
\mathcal{U}({\bf x}^{N})],
\label{eq:grandcan}
\end{equation}
where $\lambda$ is the de Broglie thermal wavelength and $\mathcal{U}$ is
the potential energy for the $N$-particle configuration ${\bf x}^{N}=(x_1,
...,x_N)$. Our simulations are performed along the following steps (for a
detailed description of the grand-canonical Monte Carlo method see
Ref.~\cite{DFrenkel}):

1. Start from an initial state with random
positions of $N$ particles;

2. A random displacement is applied to a particle selected at random. This move
is accepted with probability
\begin{equation}
\min\{1,\exp[-\beta({\mathcal U}_{\rm new}-{\mathcal U}_{\rm old})]\},
\end{equation}
where ${\mathcal U}_{\rm old}$ and ${\mathcal U}_{\rm new}$ denote, here and
in the following, the potential energy before and after the move, respectively;

3. The creation of a new particle at a random position is accepted with a
probability
\begin{equation}
\min\left\{1,\frac{L}{\lambda(N_{\rm old}+1)}
\exp[-\beta({\mathcal U}_{\rm new}-{\mathcal U}_{\rm old})]\right\},
\end{equation}
where $N_{\rm old}$ denotes, here and in the following, the particle number
before the move;

4. The removal of a randomly selected particle is accepted with a probability
\begin{equation}
\min\left\{1,\frac{\lambda N_{\rm old}}{L}
\exp[-\beta({\mathcal U}_{\rm new}-{\mathcal U}_{\rm old})]\right\};
\end{equation}

5. Repeat steps 2 to 4, for a long enough time to reach the equilibrium state.

6. Repeat steps 2 to 4, to have a sufficient number of microstates to compute
the average number of particles $\langle N \rangle$ and the density $\rho=
\langle N \rangle/L$ with good accuracy.

Note that this algorithm obeys the detailed balance principle and therefore
leads to a random sampling of the grand-canonical probability
distribution~\cite{DFrenkel}.


\begin{thebibliography}{10}
\bibitem{foot_mu}
For the case of the hard-point gas model~\cite{Benenti},
we found agreement between the results
by the two methods, provided we set $c=\mu(\rho,T)$, where
$\mu$ is the electrochemical potential obtained
from the grand-canonical Monte Carlo method. Indeed, for the hard-point
gas the dependence of $\mu$ on $\rho$ and $T$ is the same as that of the
ideal gas used as reservoir in the grand-canonical Monte Carlo method.
In general (and in particular for the case of Coulomb interaction)
the constant $c$ has not the meaning of electrochemical potential; still it
can be determined by comparison with the Monte Carlo results.
\end{thebibliography}
\end{document}